\documentclass[aps,prr,twocolumn,superscriptaddress,a4paper,floatfix]{revtex4-2}
\usepackage[utf8]{inputenc}
\usepackage{graphicx}
\usepackage{amsmath,amssymb}
\usepackage{hyperref}
\usepackage{datetime}
\usepackage{siunitx}
\usepackage{braket}
\usepackage{cleveref}
\usepackage{gensymb}
\usepackage{mathtools}

\newcommand \beq{\begin{eqnarray}}
\newcommand \eeq{\end{eqnarray}}

\newcommand \mk{\mathbf{k}}

\begin{document}

\title{Optomechanical Response of a Strongly Interacting Fermi Gas}

\author{Victor Helson}
\affiliation{Ecole Polytechnique F\'ed\'erale de Lausanne, Institute of Physics, CH-1015 Lausanne, Switzerland}
\affiliation{Center for Quantum Science and Engineering, Ecole Polytechnique F\'ed\'erale de Lausanne ,CH-1015 Lausanne, Switzerland}
\author{Timo Zwettler}
\affiliation{Ecole Polytechnique F\'ed\'erale de Lausanne, Institute of Physics, CH-1015 Lausanne, Switzerland}
\affiliation{Center for Quantum Science and Engineering, Ecole Polytechnique F\'ed\'erale de Lausanne ,CH-1015 Lausanne, Switzerland}
\author{Kevin Roux}
\affiliation{Ecole Polytechnique F\'ed\'erale de Lausanne, Institute of Physics, CH-1015 Lausanne, Switzerland}
\affiliation{Center for Quantum Science and Engineering, Ecole Polytechnique F\'ed\'erale de Lausanne ,CH-1015 Lausanne, Switzerland}
\affiliation{Université Grenoble Alpes, CEA Leti, MINATEC campus, 38054 Grenoble, France}
\author{Hideki Konishi}
\affiliation{Ecole Polytechnique F\'ed\'erale de Lausanne, Institute of Physics, CH-1015 Lausanne, Switzerland}
\affiliation{Center for Quantum Science and Engineering, Ecole Polytechnique F\'ed\'erale de Lausanne ,CH-1015 Lausanne, Switzerland}
\affiliation{Department of Physics, Graduate School of Science, Kyoto University, Kyoto 606-8502, Japan}
\author{Shun Uchino}
\affiliation{Advanced Science Research Center, Japan Atomic Energy Agency, Tokai 319-1195, Japan}
\author{Jean-Philippe Brantut}
\affiliation{Ecole Polytechnique F\'ed\'erale de Lausanne, Institute of Physics, CH-1015 Lausanne, Switzerland}
\affiliation{Center for Quantum Science and Engineering, Ecole Polytechnique F\'ed\'erale de Lausanne ,CH-1015 Lausanne, Switzerland}
\date{\pdfdate}

\begin{abstract}
We study a Fermi gas with strong, tunable interactions dispersively coupled to a high-finesse cavity.
Upon probing the system along the cavity axis, we observe a strong optomechanical Kerr nonlinearity originating from the density response of the gas to the intracavity field and measure it as a function of interaction strength.
We find that the zero-frequency density response function of the Fermi gas increases by a factor of two from the Bardeen-Cooper-Schrieffer to the Bose-Einstein condensate regime.
The results are in quantitative agreement with a theory based on operator-product expansion, expressing the density response in terms of universal functions of the interactions, the contact and the internal energy of the gas.
This provides an example of a driven-dissipative, strongly correlated system with a strong nonlinear response, opening up perspectives for the 
sensing of weak perturbations or inducing long-range interactions in Fermi gases.
\end{abstract}
\maketitle

\section{Introduction}

Cavity optomechanics allows for the sensing of mechanical displacements with ultimate sensitivity \cite{clerk_introduction_2010,aspelmeyer_cavity_2014,yu_quantum_2020}, by translating the position of an object into the resonance frequency of an optical cavity.
This framework naturally describes collective displacements of atoms within a cloud dispersively coupled to light in a high finesse cavity \cite{stamper-kurn_cavity_2014}.
The mode of the cavity singles out a particular collective mode of the atomic medium, the amplitude of which directly controls the effective cavity length.
The high sensitivity implies that upon injecting light in the cavity, the weak collective displacement induced by the photons feeds back on the cavity resonance position, yielding a Kerr nonlinearity for the cavity or equivalently an effective nonlinearity
for the atomic displacement. This hallmark of cavity optomechanics has been observed in the context of cold atoms for tightly confined, thermal clouds \cite{gupta_cavity_2007,purdy_tunable_2010} and homogeneous Bose-Einstein condensates (BECs)
\cite{brennecke_cavity_2008,ritter_dynamical_2009,kesler_optomechanical_2014}. Its counterpart for ideal Fermi gases has been predicted \cite{kanamoto_optomechanics_2010} but has yet to be observed. 

The feedback between the atomic displacement and the intracavity photons is connected by a response function, describing the ability of the gas to adapt its density to an external lattice potential.
For weakly interacting Bosons, the response is essentially that of a harmonic oscillator, with a frequency directly set by the recoil energy \cite{brennecke_cavity_2008} or the external trap \cite{gupta_cavity_2007}.
In contrast, the response of a strongly interacting quantum system depends on the complex interplay between interactions, geometry and quantum statistics. Although its direct calculation represents a theoretical challenge,
it was found in the last decade that many dynamical response functions of the strongly interacting Fermi gas have a universal character captured by Tan's relations \cite{tan_energetics_2008,tan_generalized_2008,tan_large_2008,braaten_universal_2008}
as observed with a wide variety of spectroscopic probes \cite{navon_equation_2010,stewart_verification_2010,hoinka_precise_2013,shkedrov_high-sensitivity_2018,mukherjee_spectral_2019}.

%%%%%%%%%%%%%%%%%%%%%%%%%%%%%%%%%%%%%%%%%%
% FIG1
%%%%%%%%%%%%%%%%%%%%%%%%%%%%%%%%%%%%%%%%%%

\begin{figure}
	\includegraphics[width = \columnwidth]{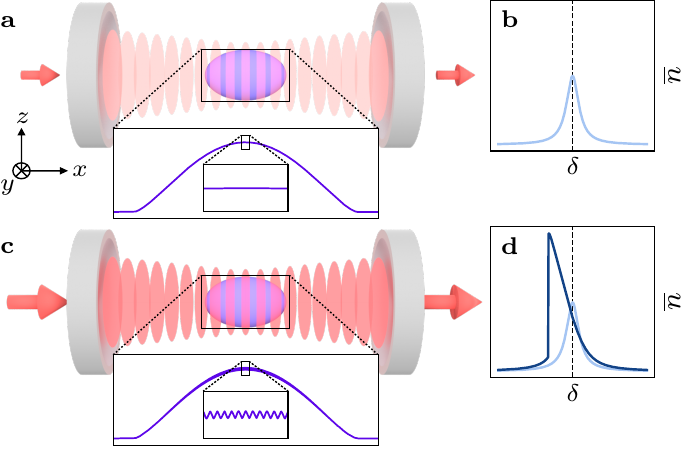}
	\caption{\textbf{Concept of the experiment} A weak laser beam dispersively interacts with a Fermi gas held within the mode of a high-finesse cavity.
	\textbf{a}~When the probe intensity is low, the dipole force exerted by the intracavity light field is negligible and the atomic density is unperturbed.
	\textbf{b}~In that case, the cavity spectrum features a symmetric, Lorentzian shape.
	\textbf{c}~For larger probe intensities, the lattice formed by the intracavity light imposes a weak static density modulation on the atoms, modifying the mode overlap with the cavity mode, thus the effective cavity length.
	\textbf{d}~The lineshape acquires an asymmetry typical of the Kerr effect. The distinctive sharp edge on the red side of the resonance is due to the onset of bistability.}
	\label{fig:Fig1}
\end{figure}

In this Letter, we report on the observation of the optomechanical Kerr nonlinearity of a degenerate, strongly interacting Fermi gas in a high-finesse cavity and measure its dependence on interaction strength in the crossover between BEC and Bardeen-Cooper-Schrieffer (BCS) regimes \cite{ketterle_making_2008,zwerger_bcs-bec_2012}.
In our experiments, the photon wavevector $k_c$ exceeds the Fermi wavevector $k_F$, so that the response function underlying the nonlinearity has a universal character inherited from the short range physics of the Fermi gas.
We use an operator product expansion (OPE) of the static response function to relate it to two thermodynamic properties of the gas, the contact and the internal energy and find that it quantitatively describes the scaling of the nonlinearity with interaction strength. 
This shows that even though the system is both strongly correlated and driven in the presence of a large feedback, its state remains controlled by a small number of universal parameters, similar to equilibrium properties.

\section{Optomechanical response}

The connection between many-body physics in the gas and the optomechanical nonlinearity originates from the structure of the dispersive light-matter coupling Hamiltonian \cite{ritsch_cold_2013,mivehvar_cavity_2021}
\begin{equation}
\hat{H}_\mathrm{lm} = \Omega \hat{a}^\dagger \hat{a}\int d^3r \hat{n}(\mathbf{r}) \cos^2 \mathbf{k}_c \mathbf{r} = \Omega \hat{a}^\dagger \hat{a} \left( \frac N 2 + \hat{M}  \right)
\label{eq:Hint}
\end{equation}
where $\Omega$ is the dispersive coupling strength, $\hat{a}$ annihilates a photon in the cavity with a mode function $\cos \mathbf{k}_c \mathbf{r}$, $\hat{n}$ is the atomic density operator and $N$ the fixed total atom number.
We have supposed that the transverse size of the cloud is much smaller than the cavity mode waist. The first part of equation~\eqref{eq:Hint} represents an average dispersive shift of the cavity while the second describes both a shift of the cavity resonance frequency originating from a collective displacement $\hat{M}$,
and an optical lattice with spacing $\pi/|\mathbf{k}_c|$ and depth $\Omega \hat{a}^\dagger \hat{a}$ imprinted onto the atoms. 

For an empty cavity or frozen atoms, $\langle \hat{M} \rangle = 0$, so that probing the cavity with vanishingly small probe power yields a symmetric, Lorentzian shaped transmission spectrum with width $\kappa$, the inverse photon lifetime, as illustrated in figures \ref{fig:Fig1}a and \ref{fig:Fig1}b.
Upon increasing probe power, the finite intracavity photon number $\bar{n} = \langle\hat{a}^\dagger \hat{a}\rangle$ imprints a density modulation on the atoms, depicted in figure \ref{fig:Fig1}c, which  yields to first order a displacement 
\begin{equation}
\langle \hat{M} \rangle = \frac{N \Omega}{8} \bar{n} \chi^R(2\mathbf{k}_c, 0),
\label{eq:M}
\end{equation}
where $\chi^R(\mathbf{q}, \omega)$ is the retarded density response function of the gas at frequency $\omega$ and wavevector $\mathbf{q}$ \cite{fetter_quantum_2003}. 
Its imaginary part is connected to the dynamical structure factor via detailed balance, and has been measured with high precision \cite{hoinka_precise_2013,carcy_contact_2019,biss_excitation_2022}.
However the zero-frequency (static) response, which is purely real, has never been measured due to the impossibility to directly observe weak, short wavelength density perturbations in a strongly interacting system. 
In the strong dispersive coupling regime $\Omega N \gg \kappa$, the cavity converts the weak, perturbative displacement into a dispersive shift with a large gain, modifying significantly the transmission of the atom-cavity system, as illustrated in figure \ref{fig:Fig1}d.

%%%%%%%%%%%%%%%%%%%%%%%%%%%%%%%%%%%%%%%%%%
% FIG2
%%%%%%%%%%%%%%%%%%%%%%%%%%%%%%%%%%%%%%%%%%

\begin{figure}
	\includegraphics[width=\columnwidth]{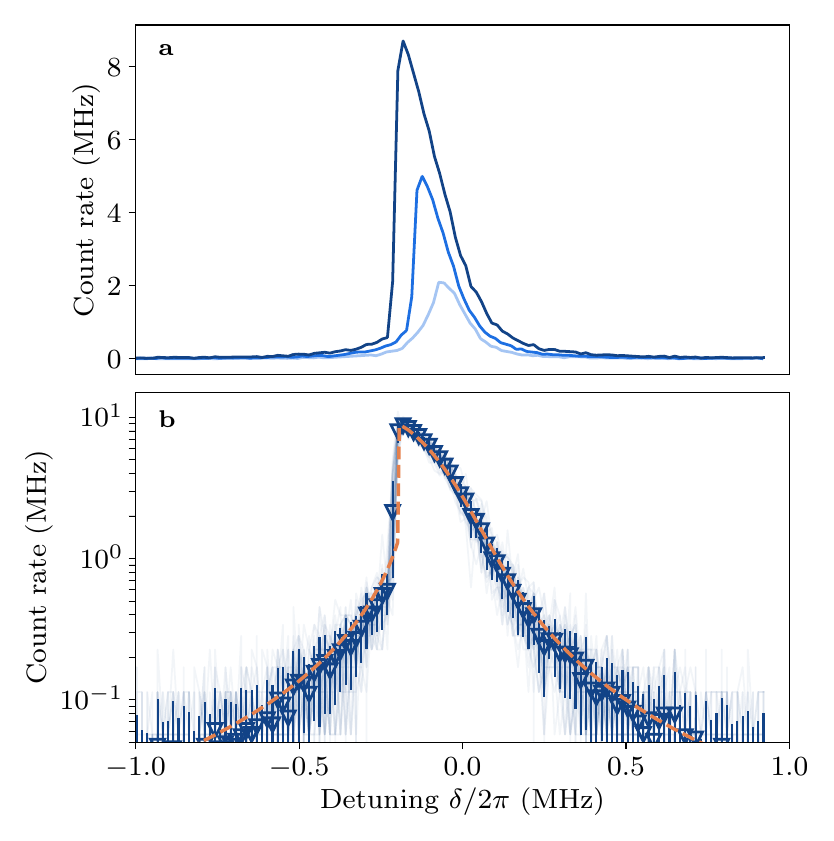}
	\caption{\textbf{Optomechanical nonlinearity for the unitary Fermi gas}.
	\textbf{a}~Transmission spectra of the cavity for increasing probe power (light blue $\bar{n}_0 = 300$, blue $\bar{n}_0 = 700$, and dark blue $\bar{n}_0 = 1300$). For weak probes we observe symmetric Lorentzian profiles (light blue).
	As the probe power increases the lineshape is distorted and features a sharp edge on the red side of the resonance (dark blue). $\delta = 0$ is the frequency of the unperturbed, dispersively shifted, atom-cavity resonance. The spectra are averaged $20$ times.
	\textbf{b}~Aggregated transmission spectra (blue triangles) and individual measurements (grey lines) taken at fixed probe intensity ($\bar{n}_0 = 1300$), in logarithmic scale. Dashed orange line: fit of equation~\eqref{eq:lineshape} to the averaged curve. The error bars represent the standard deviation over 20 realizations.}
	\label{fig:Fig2}
\end{figure}

%%%%%%%%%%%%%%%%%%%%%%%%%%%%%%%%%%%%%%%%%%

Accounting for the displacement induced by the probe, described by equation~\eqref{eq:M}, as well as the external driving and cavity dissipation in the equations of motion for the intracavity field yields a Kerr optical nonlinearity. Starting from the Hamiltonian \ref{eq:Hint} and using the input-output formalism in the frame rotating at the drive frequency, the mean-field equations of motion for the intracavity field reads
\begin{equation}
\langle \dot{\hat{a}} \rangle=-\Big(i\Delta+\frac{\kappa}{2}\Big)\langle \hat{a} \rangle -i\Omega\Big(\frac{N}{2}+\langle\hat{M}\rangle \Big)\langle \hat{a} \rangle -\sqrt{\kappa}b_{\text{in}}
\end{equation}
where $\Delta$ is the drive-cavity detuning, and $b_{\text{in}}$ is the driving amplitude. In the steady state, the atomic degrees of freedom are described by Eq~\ref{eq:M}, and the intracavity photon number reads (see Appendix \ref{app:photon_number})
\begin{equation}
	\bar{n} = \frac{\bar{n}_0}{1+\frac{4}{\kappa^2}(\delta + \eta\bar{n})^2}
	\label{eq:lineshape}
\end{equation}
where $\bar{n}_0=4 |b_{\text{in}}|^2/\kappa$ is the maximum photon number, $\delta=\Delta + \Omega N/2$ the detuning with respect to the dispersively shifted cavity resonance and $\eta= \Omega\langle \hat{M} \rangle/\bar{n} = N\Omega^2 \chi^R(2\mathbf{k}_c, 0) / 8$ measuring the strength of the Kerr nonlinearity.
This describes both the intracavity field self-amplified via the atomic medium and the atomic density fluctuations interacting via the cavity field.

\section{Experiment}
\subsection{Observation of the optomechanical response}

We perform experiments on degenerate balanced two-components Fermi gases of $^6$Li typically comprising $6 \times 10^5$ atoms, held in a crossed dipole trap with frequencies $(\omega_x, \omega_y, \omega_z) = 2\pi\times(187, 565, 594)\,$Hz along the $x$, $y$ and $z$ directions respectively.
A homogeneous magnetic field $\mathbf{B}$ oriented along the $z$ axis is tuned in the vicinity of a broad Feshbach resonance at $832\,$G, where the gas explores the BEC--BCS crossover.
The gas has a temperature of $T = 0.08(1)\,T_F$, with $T_F = \hbar (\omega_x \omega_y \omega_z 3N  )^{1/3}$ the Fermi temperature.  

The atoms are prepared in the mode of the cavity oriented along $x$ with a finesse of $47'100(700)$ and $\kappa = 2\pi \times 77(1)\,$kHz \cite{roux_cavity-assisted_2021}. We probe the system using light linearly polarized along the $z$ axis, matched to the TEM$_{00}$ mode of the cavity.
The cavity resonance frequency is detuned by $-2\pi \times 13.9\,$GHz from the D1 $\pi$ transition of $^6$Li at $832\,$G, yielding $\Omega=2\pi\times11.2$ Hz. With $\Omega N /2\kappa \approx 45$, the system operates deeply in the strong dispersive coupling regime.
The probe frequency is dynamically swept from the blue to the red side of the cavity resonance frequency over a range of $2\pi \times 3\,$MHz in $3\,$ms during a single experimental run, and the photons arrival times are recorded with a single photon counting module.
The sweep rate was chosen to be slow enough compared with the typical dynamical time scales of the system to ensure that we probe its steady state and fast enough to minimize atomic losses during the probe process.
Losses lead to a change of the dispersive coupling during the sweep representing a systematic error on the non-linearity (see Appendix \ref{app:natoms} and \ref{app:losses}).

Typical transmission lineshapes acquired for a unitary Fermi gas are depicted in figure~\ref{fig:Fig2}\textbf{a}. For low probe intensity, we record symmetric Lorentzian profiles, with a fitted linewidth of $2\pi \times 116(2)\,$kHz, enlarged compared to the inverse photon lifetime by technical fluctuations.
As the intracavity photon number is increased, the Kerr nonlinearity originating from atomic displacements distorts the profiles. For the largest probe power, we observe a distinctive sharp edge towards the red side of the cavity resonance, due to the onset of bistability predicted by equations \cite{mccall_instabilities_1974}.
The accurate determination of $\eta$, which quantifies the Kerr nonlinearity, requires a high signal-to-noise ratio that we obtain by aggregating multiple traces taken in similar experimental conditions and averaging them (see Appendix \ref{app:data_analysis}). A fit of equation~\eqref{eq:lineshape} to the averaged profiles then yields a precise value of $\eta$.
Figure~\ref{fig:Fig2}\textbf{b} displays 20 individual traces, their average and its fitted profile, showing excellent agreement with the model of equation~\eqref{eq:lineshape} over two orders of magnitude of dynamic range.

\begin{figure}
	\includegraphics[width=\columnwidth]{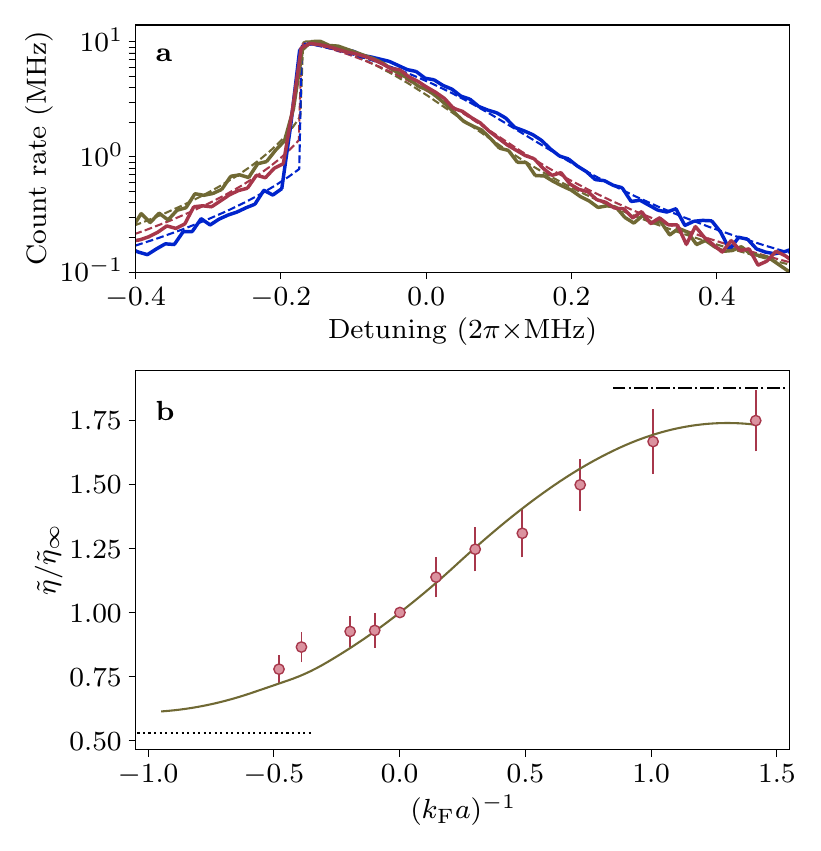}
	\caption{\textbf{Optomechanical nonlinearity in the BEC-BCS crossover}
	\textbf{a} Averaged transmission spectra and their fit using equation~\eqref{eq:lineshape}, for $1/k_Fa = -0.48, 0, 1.42$ (olive, red and blue, respectively) for fixed atom number and probe power, showing an increase of the nonlinearity with increasing $1/k_Fa$.
	\textbf{b} Density response extracted from fits to the distorted transmission profiles, normalized by the value measured at unitarity, as a function of the interaction parameter across the BEC-BCS crossover (red circles).
	The prediction of equation~\eqref{eq:OPE} is shown including the internal energy contribution (solid line), based on the contact calculated in \cite{hoinka_precise_2013}.
	The horizontal dotted and dash-dotted lines represent values of the density response computed on the BCS side obtained by cancelling the contact term of equation~\eqref{eq:OPE} and on the BEC side by considering the response of a BEC comprising $N/2$ molecules \cite{pitaevskii_bose-einstein_2016}, respectively.
	The error bars represent the uncorrelated combination of statistical fluctuations of the measurement fit uncertainties and systematic effects due to atom losses.}
	\label{fig:Fig3}
\end{figure}

\subsection{Response in the BEC-BCS crossover}

We now observe the variations of the Kerr nonlinearity, and thus of $\chi^R(2\mathbf{k}_c, 0)$ with interaction strength in the BEC--BCS crossover, by repeating this measurement at different magnetic fields around the Feshbach resonance, keeping the other conditions identical.
Typical observations are presented in figure~\ref{fig:Fig3}\textbf{a} for $B = 710, 832, 950\,$G, corresponding to interaction parameters $1/k_Fa = -0.48, 0, 1.42$, with $a$ the scattering length, together with the fit to equation~\eqref{eq:lineshape} showing a large increase of the response as the gas crosses over from the BCS to the BEC regime.  

In our regime, where $2k_c \approx 2.5 k_F$, the response function $\chi^R(2\mathbf{k}_c, 0)$ depends on the short-range physics of the gas. This can be exploited to connect the observations of Kerr nonlinearity to the universal thermodynamics of the many-body ground state of the Fermi gas.
To this end, we perform an operator product expansion (OPE) of $\chi^R(\mathbf{q}, 0)$ up to second order in $k_F/q$, leading to the expression \cite{son_short-distance_2010,goldberger_structure-function_2012,hofmann_current_2011} (see Appendix \ref{app:OPE})
\begin{equation}
	\chi^R(\mathbf{q}, 0) = -\frac{2}{\epsilon_{{q}}} - \frac{\pi\widetilde{\mathcal{I}}}{\epsilon_{{q}}q}\left[1 - \frac{8+24\pi-6\pi^2}{3\pi^2qa}\right] - \frac{8\widetilde{\mathcal{H}}}{3\epsilon^2_{{q}}}
	\label{eq:OPE}
\end{equation}
where $\epsilon_{\mathbf{q}} = \hbar^2q^2/2m$ and $m$ is the mass of $^6$Li. The expression relates $\chi^R(\mathbf{q}, 0)$ to two universal functions of the interaction parameter, the total contact $\widetilde{\mathcal{I}}$ and total internal energy $\widetilde{\mathcal{H}}$ per particle of the cloud.
A similar expression at nonzero frequency has been used to relate the dynamical and static structure factors to $\widetilde{\mathcal{I}}$ \cite{son_short-distance_2010,goldberger_structure-function_2012,hofmann_current_2011,hofmann_deep_2017}.
Compared to the structure factor results, the presence of the internal energy contribution is specific to the real part of the response function. Both $\widetilde{\mathcal{I}}$ and $\widetilde{\mathcal{H}}$ have been computed and measured accurately in the low-temperature regime in the BEC-BCS crossover \cite{kuhnle_studies_2011,kuhnle_universal_2010,stewart_verification_2010,sagi_breakdown_2015,ness_observation_2020,navon_equation_2010,horikoshi_ground-state_2017,konishi_universal_2021}.
The presence of $\widetilde{\mathcal{H}}$ in the expansion is necessary in order for the left hand side of equation~\eqref{eq:OPE} to match the response of a non-interacting Fermi gas given by Lindhard function in the BCS regime, to second order in $k_F/q$.
On the far BEC side, the OPE suffers from an intrinsic singularity at $qa=2$ (see Appendix \ref{app:OPE}). There, equation~\eqref{eq:OPE} should be interpreted as an extrapolation from the large $qa$ limit to effectively circumvent the singularity in the BEC regime.

To compare with theory, we systematically measure $\tilde{\eta} = \eta/\Omega^2N$ as a function of $B$. Here, $N$ is determined independently for each experimental realization before and after the nonlinearity measurement using non-destructive cavity-assisted methods \cite{roux_cavity-assisted_2021}.
Converting directly $\eta$ into an absolute measurement of $\chi^R(2\mathbf{k}_c, 0)$ requires the knowledge of the photon detection efficiency, which is prone to systematic errors. Instead, we normalize the results by the strength of the nonlinearity measured at unitarity $\tilde{\eta}_{\infty}$,
allowing for the precise determination of the relative variations of the response function in the BEC-BCS crossover. This amounts to using the unitary gas data to calibrate the intracavity photon number. The variations are presented in figure \ref{fig:Fig3}b, showing a smooth increase by a factor of two from the BCS to the BEC regime.

We calculate $\chi^R(\mathbf{q}, 0)$ from equation~\eqref{eq:OPE} using the values of $\widetilde{\mathcal{I}}$ for harmonically trapped gases in the BEC-BCS crossover calculated by Monte-Carlo methods \cite{hoinka_precise_2013}, and successfully compared with various experimental observations \cite{hoinka_precise_2013,kuhnle_studies_2011,konishi_universal_2021}.
We also infer from it the variations of $\widetilde{\mathcal{H}}$ using the adiabatic relation \cite{tan_large_2008} (see Appendix \ref{app:OPE}). The results are shown as a solid line in figure \ref{fig:Fig3}b. The agreement with the experimental data is good over the whole range of interaction strength, confirming the deep connection between the optical nonlinearity and the universal, many-body physics of the Fermi gas.
This adds the static optomechanical response to the set of response functions experimentally accessible following Tan's relations. The agreement with equation~\eqref{eq:OPE} indicates that the increase of nonlinearity towards the BEC regime originates predominantly from the increase of the contact.
On the BEC side, the agreement validates our extrapolation of expression~\eqref{eq:OPE}, and calls for a deeper understanding of the singularities of the OPE.

\section{Discussion and conclusion}

The amplifying ability of the cavity is striking considering that the maximum depth of the probe-induced lattice throughout the entire measurement process is only $0.2\,E_R$, with $E_R = h \times 73.7\,$kHz the recoil associated with the cavity photons, producing a density modulation with a relative depth of the order of a few percent.
The accuracy of our method is mostly limited by atomic losses in the trap, especially on the BEC side, and by probe-induced losses which limit the measurement time and usable probe strength (see Appendix \ref{app:exp_procedure}).
While losses are partially of technical origin, we also expect a fundamental contribution originating from instabilities on the red side of the cavity and amplification due to dynamical backaction on the blue side \cite{kippenberg_cavity_2007}. Including losses in the microscopic description could be performed in the framework of non-Hermitian perturbation theory \cite{pan_non-hermitian_2020}.

Compared with spectroscopic probes also sensitive to the contact, the optomechanical coupling operates in the static limit where driving the cavity changes the properties
of the steady-state of the system. In fact, the correspondence established by equation~\eqref{eq:M} also translates the optical nonlinearity into an interaction between the density fluctuations mediated by light,
repulsive on the blue side of the cavity resonance and attractive on the other, similar to cavity-induced squeezing in spin ensembles \cite{leroux_implementation_2010}. Our experiment shows that this effect is controlled by the contact,
opening fascinating perspectives for the study of the interplay between contact interactions and long-range correlations. 

Last, we could extend our optomechanical response measurement to the vicinity of a photo-association transition, where the dispersive shift acquires directly a contribution proportional to the contact \cite{konishi_universal_2021}. Correspondingly, the optomechanical response functions such as the contact-density and contact-contact responses, describing three and four-body effects, which to our knowledge have never been observed experimentally.

\section*{Acknowledgements}

We acknowledge discussions with Tobias Donner and Johannes Hofmann, and thank Joaquin Drut and Hui Hu for providing the contact data of reference \cite{hoinka_precise_2013}.
We acknowledge funding from the European Research Council (ERC) under the European Union Horizon 2020 research and innovation programme (grant agreement No 714309), the Swiss National Science Foundation (grant No 184654), EPFL, JSPS KAKENHI (grant No JP21K03436), Matsuo Foundation, and MEXT Leading Initiative for Excellent Young Researchers.

\appendix

\section{Photon number expression}
\label{app:photon_number}

We consider the coupling of a Fermi gas to light in an optical cavity in the dispersive limit, where
the light-matter coupling is given as equation (1) in the main text. In the presence of such a coupling, the Heisenberg-Langevin (input-output) equation for the photon field in the frame rotating at a driving frequency is given by \cite{ritsch_cold_2013,mivehvar_cavity_2021}
\beq
\dot{\hat{a}}=-\Big(i\Delta+\frac{\kappa}{2}\Big)\hat{a}-i\Omega\Big(\frac{\hat{N}}{2}+\hat{M} \Big)\hat{a}-\sqrt{\kappa}\hat{b}_{\text{in}},
\eeq
where $\Delta$ is the detuning with respect to the empty cavity, 
$\hat{b}_{\text{in}}$ the input field, and $\kappa$ the cavity decay rate.
We then consider the case of the coherent-state input where the coherent state description for the photon field
is reasonable and the corresponding $c$-number field $a$  obeys the following equation \cite{ritsch_cold_2013,mivehvar_cavity_2021}
\beq
\dot{a}=-\Big(i\Delta+\frac{\kappa}{2}\Big)a-i\Omega\Big(\frac{N}{2}+\langle\hat{M}\rangle \Big)a-\sqrt{\kappa}b_{\text{in}}.
\label{eq:coherent}
\eeq
Here we note that the average $\langle \hat{M}\rangle$ contains the light-matter coupling Hamiltonian $\hat{H}_{\text{lm}}$.
In the small light-matter coupling limit where the linear response analysis is allowed, 
we obtain
\beq
\langle\hat{M}\rangle\approx\frac{N\Omega }{8}\int_{-\infty}^{\infty} dt'\chi^R(2\mathbf{k}_c,t-t')\bar{n}(t').
\label{eq:density}
\eeq
Here we define the retarded density response function per atom, 
\beq
\chi^R(\mathbf{q},t)=\frac{1}{N}\int d^3R\chi^R(\mathbf{R},\mathbf{q},t),
\eeq
which can be written as the spatial integral of the following local retarded density response function:
\beq
\chi^R(\mathbf{R},\mathbf{q},t)&&=-i\theta(t)\int d^3r e^{-i\mathbf{q\cdot r}} \times \nonumber\\ &&\langle[\hat{n}(\mathbf{R+r/2}),\hat{n}(\mathbf{R-r/2}) ]\rangle_0,
\label{eq:local-density}
\eeq
with $\hat{n}$ the density operator of the gas, and $\langle\cdots\rangle_0$ denoting the average in the absence of light-matter coupling.
In order to obtain equation~\eqref{eq:density}, we used the local density approximation where
the system is locally uniform with translational and inversion symmetries~\footnote{To be precise, the local translational symmetry
ensures that the local density response function is expressed with the Fourier component of the relative coordinate~\eqref{eq:local-density},
and the local inversion symmetry ensures $\chi^R(\mathbf{R},-2\mathbf{k}_c,t)=\chi^R(\mathbf{R},2\mathbf{k}_c,t)$.}.

We now assume the steady-state solution of $a$ where time dependence of the cavity field can be neglected.
In this case, $\langle\hat{M}\rangle$ is reduced to
\beq
\langle\hat{M}\rangle=\frac{N\Omega}{8}\bar{n}\chi^R(2\mathbf{k}_c,0),
\label{eq:m}
\eeq
where $\chi^R(2\mathbf{k}_c,0)=\chi^R(2\mathbf{k}_c,\omega=0)$.
By substituting \eqref{eq:m} into \eqref{eq:coherent} under $\dot{a}=0$, we have
\beq
a=\frac{-2b_{\text{in}}/\sqrt{\kappa}}{1+\frac{2i}{\kappa}(\delta+\eta\bar{n})}.
\label{eq:a}
\eeq
By using above and the coherent state property $\bar{n}=|a|^2$,
equation~(4) in the main text is obtained.

\begin{widetext}
\section{Asymptotic retarded response function expression from the operator product expansion}
\label{app:OPE}
We discuss the asymptotic form of the density response function in terms of the operator product expansion (OPE).
Since the detailed analysis on corresponding time-ordered Green's function has already been done in Refs.~\cite{son_short-distance_2010,hofmann_current_2011,goldberger_structure-function_2012}, here we focus on the essential idea and result.

The OPE is a standard approach of quantum field theory which states that 
the product of local operators at different points in space and time can be expanded in terms of
local operators. In the case of the density response function, we can consider the following relation:

\beq
\hat{n}(\mathbf{R+r}/2,t)\hat{n}(\mathbf{R-r}/2,0)=\sum_{m}c_m(\mathbf{r},t) \hat{{\cal O}}^m(\mathbf{R},0),
\label{eq:ope}
\eeq
where $\hat{{\cal O}}^m$ is a local operator and $c_m(\mathbf{r},t)$ is a function of the relative coordinates.
For generic space-time points, the above operator identity may be useless, 
since the sum in the right hand side is taken over infinitely many local operators.
In the short range regime that is shorter than mean particle distance ($\sim1/k_F$) 
and inverse of the thermal de Broglie length
but is longer than a cutoff length of the effective theory (e.g. van der Waals length in cold gases), however, the OPE becomes a powerful tool in that
first few terms with low scaling dimensions in the operator sum dominate the identity~\cite{zwerger_bcs-bec_2012}.

In order to proceed the OPE calculation in terms of quantum field theory, it is convenient to work with the following time-ordered Green's function
\beq
\chi^T(\mathbf{q},\omega)=\frac{-i}{N}\int d^3R \int d^3r\int dt e^{i\omega t-i\mathbf{q\cdot r}} \langle T[n(\mathbf{R+r/2},t)n(\mathbf{R-r/2},0)] \rangle_0,
\eeq
where $T$ denotes time ordering. By using equation~\eqref{eq:ope}, the time-ordered function obeys the following relation:
\beq
\chi^T(\mathbf{q},\omega)=\sum_mc_m(\mathbf{q},\omega)\int d^3R \frac{\langle \hat{{\cal O}}^m(\mathbf{R})\rangle_0}{N}.
\eeq 
As defined before, $\langle\cdots\rangle_0$ is the average without $\hat{H}_{\text{lm}}$, and 
therefore $\langle \hat{{\cal O}}^m\rangle_0$ is essentially the equilibrium average of a Fermi gas.
By using the Feynman diagram technique, we can determine the so-called Wilson coefficient $c_m$ and $\chi^T$.
Up to $(k_F/q)^2$, $\chi^T$ has been obtained as follows~\cite{hofmann_current_2011,goldberger_structure-function_2012}:
\beq
\chi^T(\mathbf{q},\omega)&&\approx N c_n(\mathbf{q},\omega)+c_{{\cal I}}(\mathbf{q},\omega) \tilde{{\cal I}}+c_{{\cal H}}(\mathbf{q},\omega)
\tilde{{\cal H}}.
\label{eq:density-ope}
\eeq
Here we define 
\beq
c_n(\mathbf{q},\omega)&&=\frac{2\epsilon_q}{\omega^2-\epsilon_q^2},\\
c_{{\cal H}}(\mathbf{q},\omega)&&=\frac{4\epsilon_q}{3}\Big[\frac{1}{(\omega-\epsilon_q)^3}-\frac{1}{(\omega+\epsilon_q)^3} \Big],\\
c_{{\cal I}}(\mathbf{q},\omega)&&=-\Big[A(\mathbf{q},\omega)\Big\{I_1(\mathbf{q},\omega)+\frac{2}{A(0)}\frac{1}{\omega-\epsilon_q} \Big\}^2
+A(-\mathbf{q},-\omega)\Big\{I_1(-\mathbf{q},-\omega)+\frac{2}{A(0)}\frac{1}{-\omega-\epsilon_{-q}} \Big\}^2\nonumber\\
&&-\frac{(I_2(\mathbf{q},\omega)+I_3(\mathbf{q},\omega))}{2}-\frac{(I_2(-\mathbf{q},-\omega)+I_3(-\mathbf{q},-\omega))}{2}-\frac{4}{A(0)}\frac{1}{\omega^2-\epsilon^2_q}\nonumber\\
&& -\frac{2}{A(0)}\Big\{ \frac{1}{(\omega-\epsilon_q)^2} +\frac{1}{(\omega+\epsilon_q)^2}\Big\}-\frac{4\epsilon_q}{3A(0)}\Big\{\frac{1}{(\omega-\epsilon_q)^3}-\frac{1}{(\omega+\epsilon_q)^3}
\Big\}\Big],
\eeq
with $\epsilon_q=q^2/(2m)$, $A(0)=-\frac{4\pi a}{m}$,
\beq
A(\mathbf{q},\omega)&&=\frac{\frac{4\pi}{m}}{-\frac{1}{a}+\sqrt{-m(\omega-q^2/(4m))-i0^+ }},\\
I_1(\mathbf{q},\omega)&&=\int\frac{d^3k}{(2\pi)^3}\frac{1}{\epsilon_{\mk}}\frac{1}{\omega-\epsilon_{\mk}-\epsilon_{\mk+\mathbf{q}}+i0^+},\\
I_2(\mathbf{q},\omega)&&=\int\frac{d^3k}{(2\pi)^3}\frac{1}{\epsilon^2_{\mk}}\Big[\frac{1}{\omega-\epsilon_{\mk}-\epsilon_{\mk+\mathbf{q}}+i0^+}
-\frac{1}{\omega-\epsilon_{\mathbf{q}}+i0^+}\Big],\\
I_3(\mathbf{q},\omega)&&=\int\frac{d^3k}{(2\pi)^3}\frac{1}{\epsilon_{\mk}\epsilon_{\mk+\mathbf{q}}}\frac{1}{\omega-\epsilon_{\mk}-\epsilon_{\mk+\mathbf{q}}+i0^+}.
\eeq

We now apply the above result for the retarded density response function.
To this end, we focus on the spectral representation of Green's function~\cite{fetter_quantum_2003}, 
which allows to relate
$\chi^T$ to $\chi^R$.
By using this, it is straightforward to show
\beq
\text{Re}[\chi^R(\mathbf{q},\omega)]=\text{Re}[\chi^T(\mathbf{q},\omega)],\\
\chi^R(\mathbf{q},0)=\text{Re}[\chi^R(\mathbf{q},0)].
\eeq
Thus, it turns out that $\chi^R(\mathbf{q},0)$ can directly be determined from equation~\eqref{eq:density-ope}.
In order to obtain a simpler expression, we finally perform integrals in $I_1$, $I_2$, and $I_3$.
First, by using
\beq
&&\frac{1}{AB}=\int_0^1dx\frac{1}{[Ax+B(1-x)]^2},\\
&&\int\frac{d^3k}{(2\pi)^2}\frac{1}{[k^2+\alpha^2-i0^+]^2}=\frac{1}{8\pi\alpha},
\eeq
we can show
\beq
I_1(\pm\mathbf{q},0)=-\frac{m^2}{4q}.
\eeq
In addition,
\beq
I_2(\pm\mathbf{q},0)=-\frac{2m^3}{\pi q^3},
\eeq
can be shown by using
\beq
\frac{1}{AB^2}=2\int_0^1\frac{1-x}{[Ax+(1-x)B]^3},\\
\int\frac{d^3k}{(2\pi)^2}\frac{1}{[k^2+\alpha^2-i0^+]^3}=\frac{1}{32\pi\alpha^3}.
\eeq
Finally, by using
\beq
&&\frac{1}{ABC}=2\int_0^1dx\int_0^{1-x}dy\frac{1}{[Ax+By+C(1-x-y)]^3},\\
&&\int_0^1dx\int_0^{1-x}dy\frac{1}{\{1-(x-y)^2 \}^{3/2}}=1,
\eeq
we obtain
\beq
I_3(\pm\mathbf{q},0)=-\frac{2m^3}{\pi q^3}.
\eeq

And we also obtain 
\beq
A(\pm\mathbf{q},0) = \frac{8\pi}{mq \left(1-\frac{2}{qa} \right)},
\eeq

such that the OPE expansion for $\chi^R(\mathbf{q},0)$ reads:
\beq \label{eq:response_without_approx}
\chi^R(\mathbf{q},0)=-\frac{2N}{\epsilon_q}-\frac{8\tilde{\cal H}}{3\epsilon_q^2} -\frac{\pi \tilde{\cal I}}{2\epsilon_qq}\Big[\frac{1}{1-\frac{2}{qa}}-\frac{8}{\pi qa(1-\frac{2}{qa})} +\frac{16}{\pi^2(qa)^2 (1-\frac{2}{qa})}-\frac{8}{3\pi^2qa}  \Big].
\eeq
This expression diverges for $qa=2$, such that in the BEC regime it becomes uncontrolled. However, close to the unitary regime $qa\gg1$, the expression above reduces to equation (4) in the main text 
\beq \label{eq:response_with_approx}
\chi^R(\mathbf{q},0)=-\frac{2N}{\epsilon_q}-\frac{8\tilde{\cal H}}{3\epsilon_q^2} -\frac{\pi \tilde{\cal I}}{2\epsilon_qq}\Big[1-\frac{8+24\pi-6\pi^2}{3\pi^2qa}\Big].
\eeq
which is regular in the BEC regime. The OPE result with and without the large $qa$ limits are presented in figure~\ref{fig:figOPE}.
\end{widetext}

\begin{figure}[ht!] 
    \includegraphics{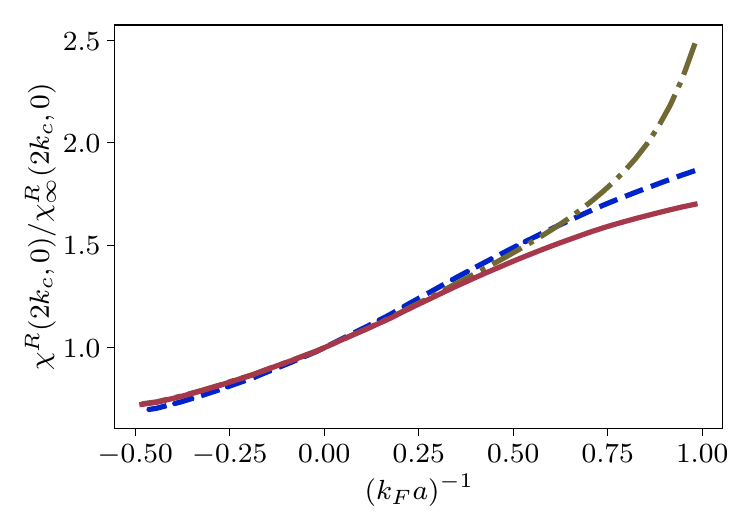}
    \caption{Comparison of different OPE expressions for $\chi^R(\mathbf{q},0)$ as a function of $(k_F a)^{-1}$ for a trap-averaged system of $6 \times 10^5$ atoms and other parameters as in the main text.
    The OPE expressions show all very similar behaviour in the unitary regime. However, in the BEC regime the expression without $qa\gg1$ approximation (\ref{eq:response_without_approx}) shows a divergence approaching $qa=2$ (olive dash dotted line),
    whereas the OPE expression with $qa\gg1$ approximation (\ref{eq:response_with_approx}) (red solid line) circumvents the singular behaviour and should be interpreted as an extrapolation from the large $qa$ limit. For comparison, also (\ref{eq:response_with_approx})
    without the internal energy contribution is shown (blue dashed line).}
    \label{fig:figOPE}
\end{figure}

\section{Experimental procedure}
\label{app:exp_procedure}
We prepare a quantum degenerate, strongly interacting Fermi gas of ${}^6 \mathrm{Li}$
following the method described in \cite{roux_cavity-assisted_2021}. At the end of the all-optical preparation, we perform transmission spectroscopy
while holding the Fermi gas in a crossed optical dipole trap with trapping frequencies of $2 \pi \times (187, 565, 594)\,$Hz along the longitudinal and transversal directions of the cavity.

The spectroscopic measurement consists of four probe pulses with $\pi$ polarization, during which we sweep the frequency of the probe beam across the cavity resonance.
First, we perform a pulse with fast sweep rate to determine the resonance of the dispersively shifted cavity, allowing us to infer atom number in a nondestructive way \cite{roux_cavity-assisted_2021}.
This is followed by a second pulse to measure the optomechanical response of the Fermi gas as described in the main text.
The third and fourth pulses measure the dispersively shifted cavity resonance after the response measurement to infer losses and the cavity resonance without atoms, respectively.

The probe power is optimized for the highest signal-to-noise ratio and is limited by the dynamic range of the single photon counter.
At high photon flux the detector response becomes nonlinear, which introduces a bias in the Kerr effect determination.
By recording the position of the cavity resonance without atoms, we can correct for slow drifts in the experimental lock chain and exactly determine the dispersive shifts.

\section{Atom number determination and losses}
\label{app:natoms}
The atom number is extracted from the dispersive shift of the cavity resonance with respect to the bare cavity
$\delta_{\mathrm{at}} = \frac{N \Omega}{2}$ with light shift $\Omega = \frac{g_0^2}{\Delta_{\mathrm{a}}}$, atom number $N$, single photon-atom coupling strength $g_0$ and atom-cavity detuning $\Delta_a$.
We take D1 and D2 $\pi$ transitions into account to infer the total atom number

\begin{equation}
    N = \frac{2\delta_{\mathrm{at}}}{g^2_{\mathrm{D1} \pi}/\Delta_{\mathrm{D1} \pi}+g^2_{\mathrm{D2} \pi}/\Delta_{\mathrm{D2} \pi}}
\end{equation}
with the coupling strengths $g_{\mathrm{D1}\pi} = 2 \pi \times 276\,$kHz, $g_{\mathrm{D2}\pi} = 2 \pi \times 391\,$kHz and 
the detunings of the cavity resonance from the respective atomic transitions at $832\,$G of $\Delta_{\mathrm{D1} \pi} = -2\pi \times 13.9\,$GHz and $\Delta_{\mathrm{D2} \pi} = -2\pi \times 23.6\,$GHz.

\begin{figure}[h!]    
    \includegraphics{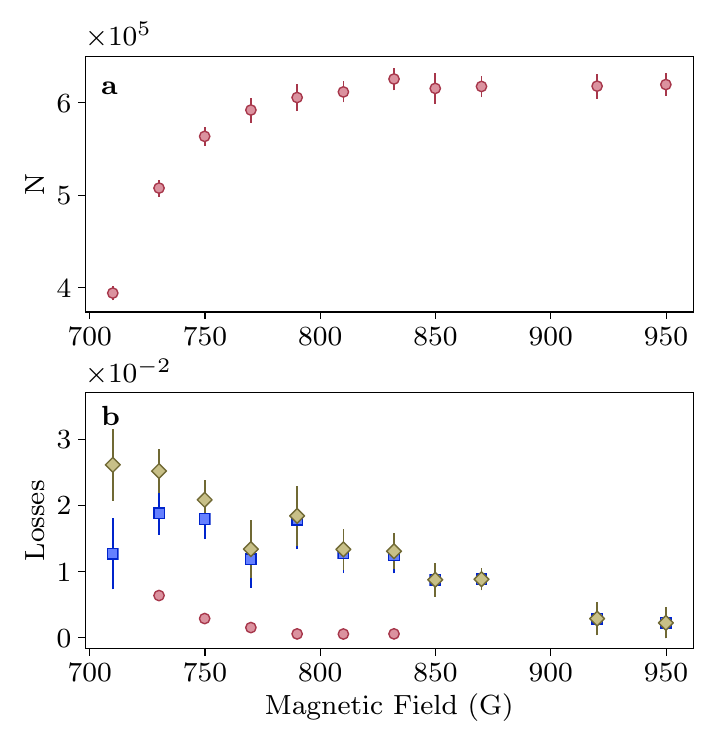}
    \caption{\textbf{a} Total atom number measured before the nonlinearity measurement, as a function of magnetic field. 
    \textbf{b} Total losses during the nonlinearity measurement as a function of magnetic field (olive diamonds) compared to trap losses due to the finite atomic lifetime (red circles)
    and probe induced losses (blue squares).}
    \label{fig:figLossesB}
\end{figure}

\begin{figure}[h!]    
    \includegraphics{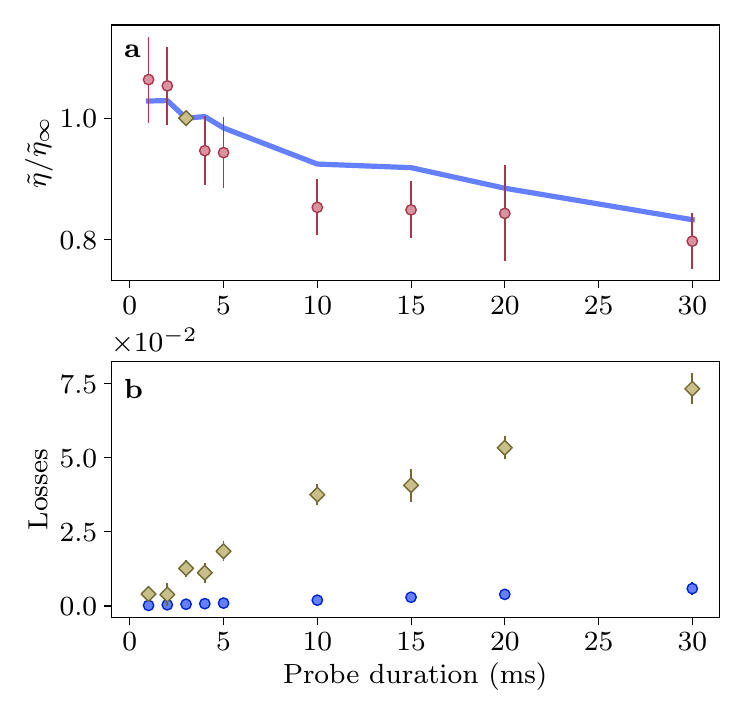}
    \caption{\textbf{a} Static response per atom $\tilde{\eta}$ as a function of the probe duration over a frequency range of $2\pi \times 2.7\,$MHz for a unitary Fermi gas,
    all other parameters identical to the main text. $\tilde{\eta}$ is normalized to the response per atom $\tilde{\eta}_{\infty}$ at $3\,$ms probe duration used in the experiment at $832\,$G (olive diamond).
    Simulation of the effect of losses on the fitted static response (light blue line).
    \textbf{b} Total atomic losses as a function of the sweep duration inferred from two dispersive shift measurements respectively performed before and after the slow sweep (olive diamonds).
    The chosen probe duration of $3\,$ms minimizes measurement-induced losses to below $1.4\%$, which would systematically shift down the static response of the gas for longer sweep durations.
    The effect of probe induced losses becomes more significant for longer probe durations, whereas the trap losses remain small over the time of probing (blue circles).}

    \label{fig:losses_vs_sweep_dur}
\end{figure}

Using a fast and non-destructive measurement of the dispersive shift, we determine the atom number before and after the optomechanical response measurement and infer the atomic losses encountered during the measurement. We investigated the losses as a function of the optomechanical response measurement duration, as shown in figure~\ref{fig:losses_vs_sweep_dur}b for the unitary Fermi gas. For durations shorter than $5\,$ms, the losses are below $2\%$, while reaching up to $7.2\%$ for a sweep duration of $30\,$ms.

Losses originate both from the finite lifetime of the gas in the trap and from the measurement itself. To disambiguate between these two effects, we measure the losses due to the finite lifetime of the gas, depending on interaction strength, by replacing the slow optomechanical response measurement between the two dispersive shift measurements by another dispersive shift measurement. This also allows us to infer the exact atom number at the position of the slow sweep.
The results for the losses corresponding to an optomechanical response measurement duration of $3\,$ms are presented in figure~\ref{fig:figLossesB}, which shows a strong
increase of the loss rate towards the BEC side due to three-body losses \cite{ketterle_making_2008}. We find that the total losses at $832\,$G are below $2\%$, matching the value measured for figure~\ref{fig:losses_vs_sweep_dur}b. Therefore, for durations of $3\,$ms, trap losses have a small contribution for a unitary gas, and we infer measurement-induced losses of $1.3\%$. On the BEC side, we also infer measurement-induced losses below $2\%$ for $3\,$ms duration, hence the choice for the main text data presented in figure~\ref{fig:losses_vs_sweep_dur}a. This ensures that the measurement-induced losses do not shift the cavity resonance by more than its linewidth while maintaining a high signal-to-noise ratio of the photon count rate at all detunings.

Note that the equilibrium condition for the atomic gas is easily fulfilled since we increase the probe power from the background to its maximal value in about $1\,$ms, which is long compared to the dynamical time scale of atomic motion given by the recoil time of $13.57\,\mu$s. 

Even though losses are small compared with the total atom number, the change of dispersive shift during the measurement yields a systematic decrease of the observed non-linearity, which we account for in the error bars (see below).

\section{Data analysis}
\label{app:data_analysis}
\begin{figure}[h!] 
    \includegraphics{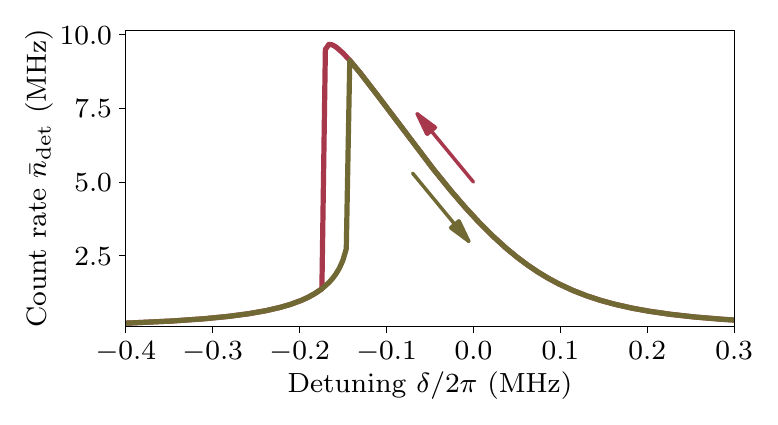}
    \caption{Solution of the cubic polynomial given by equation~\eqref{eq:dat_ana_2} for the fitted parameters at $832$\,G.
    The two stable solutions of the cubic polynomial, which are explored by sweeping the frequency from the blue to the red side (red)
    and from the red to the blue side (olive). A sweep direction downward in frequency is chosen in the experiment to avoid measurement-induced heating resulting from a sudden increase in intracavity photon number.}
    \label{fig:fig1}
\end{figure}

The photon count rate $\bar{n}_{\mathrm{det}}$ recorded from transmission spectroscopy is related to the intracavity photon number $\bar{n}$ by

\begin{equation} \label{eq:dat_ana_1}
\bar{n}_{\mathrm{det}} = \epsilon \frac{\kappa}{2} \bar{n} 
\end{equation}
with photon-detection efficiency $\epsilon$ and cavity intensity decay rate $\kappa$. Inserting equation~\eqref{eq:dat_ana_1} into equation~\eqref{eq:a} yields the model for the Kerr nonlinear resonator

\begin{equation} \label{eq:dat_ana_2}
    \bar{n}_{\mathrm{det}} = \frac{\bar{n}_{0}^{\mathrm{det}}}{1+(2 \frac{\delta}{\kappa} + 4 \frac{\eta^{\prime}}{\kappa^2} \bar{n}_{\mathrm{det}})^2}
\end{equation}
with the maximal detected photon count rate $\bar{n}_{0}^{\mathrm{det}} = \epsilon \frac{\kappa}{2} \bar{n}_{0}$ and Kerr non-linearity $\eta^{\prime} = \frac{\eta}{\epsilon}$. Equation~\eqref{eq:dat_ana_2} reduces to a third order polynomial in $\bar{n}_{\mathrm{det}}$, which can be solved for each value of $\delta$.
Figure~\ref{fig:fig1} displays the two stable solutions of the equation for a set of parameters reproducing the experimental conditions. In the bistable regime the equation has three real solutions, of which the two shown solutions correspond to the stable branches of the model.
We pick the largest of the two real stable solutions which corresponds to our procedure of downwards sweeping the probe frequency. That branch is depicted by the red curve of figure~\ref{fig:fig1}. The downward sweep direction is chosen to avoid measurement-induced heating due to a sudden increase in intracavity photon number.

The numerical solution is then used as a fit function for $\bar{n}_{\mathrm{det}}$ as a function of $\delta$. We acquire $30$ transmission spectra obtained with the same experimental parameters on different realizations of the atomic samples. We overlap the spectra on their sharp edge before averaging them together and fitting the averaged profile on equation~\eqref{eq:dat_ana_2} to extract $\eta$, as depicted by figure 2b of the main text.
We found that the fitted value of the nonlinearity parameter $\eta$ is sensitive to the shape of the tails of the transmission profile. To obtain a good agreement between the fit and the averaged profiles, we fit the logarithm of the latter, thus equally weighting data points at the peak and in the wings of the profile. This results in the typical data shown in figure 2b of the main text, and their fit with good agreement over 2 orders of magnitude.
In the procedure, we fix $\bar{n}_{0}^{\mathrm{det}}$ to the maximal detected number of photons. Moreover, the effective cavity linewidth is increased compared to the fast sweep configuration due to technical fluctuations at the slow sweep rate at which the experiment is operated. Measuring the linewidth of the cavity with slow sweep rates, we obtain $\kappa = 2\pi \times 116(2)\,$kHz, which we also set as fixed in the fitting procedures.

The statistical errors are estimated using a bootstrap method. We randomly resample the 30 profiles taken for one set of experimental conditions to produce 20 datasets, which are then averaged and fit, yielding 20 statistical realizations of $\eta$ for the initial 30 spectra. These realizations are averaged, their mean used as the measured value of $\eta$ in the main text, and their standard deviations used as statistical error estimate.
These are combined with estimates for the systematic error due to atom losses (see below) to produce the error bars.

\section{Systematic error due to losses}
\label{app:losses}
In figure~\ref{fig:losses_vs_sweep_dur}a, we display the effect of losses on the fitted static response. By assuming a constant atom loss rate $L$
over the probe duration for the trap and probe-induced losses as observed in figure~\ref{fig:losses_vs_sweep_dur}b, we simulate the distortion of the Kerr nonlinear profile by modifying equation~\eqref{eq:dat_ana_2} to
\begin{align} \label{eq:loss_sim}
    \eta^{\prime}(t) &= \tilde{\eta}^{\prime} \Omega^2 (N-Lt)\\
    \delta(t) &= \frac{d\Delta}{dt} t + \Omega/2 (N-Lt)
\end{align}
with the detuning from the bare cavity resonance $\Delta$.
We then fit the simulated, distorted profiles using equation~\eqref{eq:dat_ana_2} and use the result to estimate the effect of losses on the static response (light blue line figure~\ref{fig:losses_vs_sweep_dur}a).
As expected, we observe that losses decrease the fitted static response, due to the variations of the dispersive shift during the measurement. The result of the loss simulation reproduces well the tendency of the fitted static response, as can be seen in figure~\ref{fig:losses_vs_sweep_dur}a.
For the chosen probe duration of $3\,$ms, we perform the simulation for the losses at each magnetic field. %We observe that the loss effect on $\kappa$ is below $3\%$ and fix it in the fit.
Using this simulation, we estimate the systematic error of the fitted $\eta^{\prime}$ to be $4\%$ compared to its undistorted value. This is now included in the error bar for the data shown in the main text, through independent error combination with the statistical fluctuations inferred from the bootstrap method (see above).

%\bibliography{optomechanics}

\end{document}